\documentclass[%
 reprint,
superscriptaddress,
 amsmath,amssymb,
 aps,
]{revtex4-2}

\usepackage{graphicx}
\usepackage{dcolumn}
\usepackage{bm}
\usepackage{svg}
\usepackage{comment}
\usepackage{xcolor}

\usepackage{comment}
\begin{document}

\preprint{APS/123-QED}

\title{Short Bragg pulse spectroscopy for a paraxial fluids of light}

\author{Clara Piekarski}
\affiliation{Laboratoire Kastler Brossel, Sorbonne Université, CNRS, ENS-PSL Research University, Coll\`ege de France, Paris 75005, France}

\author{Wei Liu}%
\affiliation{Laboratoire Kastler Brossel, Sorbonne Université, CNRS, ENS-PSL Research University, Coll\`ege de France, Paris 75005, France}

\author{Jeff Steinhauer}
\affiliation{Laboratoire Kastler Brossel, Sorbonne Université, CNRS, ENS-PSL Research University, Coll\`ege de France, Paris 75005, France}
\affiliation{Department of Physics, Technion—Israel Institute of Technology, Technion City, Haifa 32000, Israel}

\author{Elisabeth Giacobino}
\affiliation{Laboratoire Kastler Brossel, Sorbonne Université, CNRS, ENS-PSL Research University, Coll\`ege de France, Paris 75005, France}

\author{Alberto Bramati}
\affiliation{Laboratoire Kastler Brossel, Sorbonne Université, CNRS, ENS-PSL Research University, Coll\`ege de France, Paris 75005, France}

\author{Quentin Glorieux}
 \email{quentin.glorieux@lkb.upmc.fr}
\affiliation{Laboratoire Kastler Brossel, Sorbonne Université, CNRS, ENS-PSL Research University, Coll\`ege de France, Paris 75005, France}

\date{\today}

\begin{abstract}
We implement Bragg spectroscopy in a paraxial fluid of light.
Analogues of short Bragg pulses are imprinted on a photon fluid by wavefront shaping using a spatial light modulator.  
We measure the dispersion relation and evidence a parabolic single-particle regime as well as a linear phonon regime even for very weakly interacting photons and low sound velocity. 
Finally, we report a measurement of the static structure factor, $S(k)$, and we demonstrate the presence of pair-correlated excitations, revealing  indirectly  the  quantum  depletion in  a paraxial fluid of light 

\end{abstract}

\maketitle


Fluids of light in the paraxial configuration have  emerged as an original approach to study degenerate Bose gases \cite{carusotto2014superfluid}. 
Several important results have recently established this platform as a potential analogue quantum simulator, including the demonstrations of superfluidity of light \cite{fontaine2018,michel2018superfluid}, the observation of the Berezinskii–Kosterlitz–Thouless transition \cite{situ2020dynamics} and pre-condensation \cite{vsantic2018nonequilibrium}, the evidence of photon droplets \cite{wilson2018observation} and the creation of analogue rotating black hole geometries \cite{vocke2018rotating,jacquet2020polariton}.
Paraxial fluids of light rely on the direct mathematical analogy that can be drawn between the Gross-Pitaevskii equation describing the mean field evolution of a Bose-Einstein condensate (BEC) and the non-linear Maxwell equation describing the propagation of light within a $\chi^{(3)}$ non-linear medium \cite{frisch1992transition,carusotto2014superfluid}. 
The growing interest about this platform comes from the various advantages that makes paraxial fluids of light a complementary system to atomic BEC.
First, optical detection techniques are highly sensitive (single-photon counting, homodyne detection) and allows for measuring with high precision the density distribution in position and momentum space as well as the phase.
Second, since there is no gravity force acting on a photon fluid, there is no need for a trapping potential and homogeneous density can be easily achieved.
Moreover, an external control potential can be applied either during the entire evolution \cite{michel2018superfluid,fontaine2019attenuation}, or only for a short period of time as in this work.

An essential characterization tool for atomic BEC is coherent Bragg spectroscopy \cite{blakie2002theory}.
The original implementation was based on two-photon Bragg scattering and allowed for the measurement of the true momentum distribution, which  was  significantly  narrower  than  that observed by time of flight \cite{PhysRevLett.82.4569}.
For atomic BEC, it has been realized very early that this technique would also allow for measuring the dispersion relation \cite{PhysRevLett.83.2876}, which describes how each frequency component of a wavepacket evolves and the dynamic structure factor, which is the Fourier transform of the density correlation function \cite{PhysRevLett.88.120407} and is essential to the description of many-body systems \cite{nozieres2018theory,griffin1993excitations}.

Several variants of this method have been developed for exciton-polaritons \cite{PhysRevLett.106.255302} and for atomic BEC, including momentum-resolved spectroscopy \cite{ernst2010probing}, multi-branch spectroscopy \cite{PhysRevLett.90.060404} and tomographic imaging \cite{PhysRevLett.88.220401}.
Most of these techniques rely on measuring the energy of the condensate's linear excitations known as Bogoliubov quasi-particles.
Interestingly, in paraxial fluids of light, the dispersion relation has been recently obtained  not using Bragg spectroscopy but by measuring the group velocity of two counter-propagating wavepackets in the transverse plane \cite{fontaine2018}.
However, this experiment, as well as a similar implementation in a thermo-optic medium \cite{vocke2015experimental}, have a resolution for weak non-linearities (small sound velocities).
Indeed, in this regime, the two wavepackets propagate too slowly to separate and due to interferences no analytical  dispersion relation can be derived \cite{fontaine2020interferences}. 

In this Letter, we implement an optical analogue of Bragg spectroscopy to measure the phonon dispersion and the static structure factor in a paraxial fluid of light.
We show that short Bragg pulses used for phase imprinting technique in atomic BEC can be achieved in fluids of light using wavefront shaping with a spatial light modulator.
This technique not only allows for measuring the dispersion relation with a better resolution at small sound velocities but it also gives access to the excitation strength for phonons: the static structure factor.
We found that the static structure factor is significantly reduced from that of free particles, revealing indirectly the quantum depletion, consisting of pair-correlated particles, in a paraxial fluid of light \cite{PhysRevLett.83.2876}.

This paper is organized as follows. 
We first introduce the formalism of a paraxial fluid of light and the short Bragg pulse technique which inspired our approach.
We describe numerically and experimentally the optical implementation of Bragg spectroscopy.
We then measure the dispersion relation and evaluate the maximum resolution.
Finally, we present a measurement of the zero temperature static structure factor in agreement with the Feynman relation for an homogeneous Bose gas \cite{feynman1954atomic,nozieres2018theory}.



In a third-order nonlinear Kerr medium, the evolution of the electric field is given by the nonlinear Schrödinger equation (NLSE), written within the paraxial and slowly-varying-envelope approximation as:
\begin{equation}
    i\dfrac{\partial E}{\partial z}= \left(-\dfrac{1}{2k_0}\nabla_\perp^2-k_0n_2|E|^2+i\dfrac{\alpha}{2}\right)E \text{,}
    \label{nlse}
\end{equation}
where $k_0$ is the wavevector, $\alpha$ is the linear absorption coefficient, $n_2$ is the nonlinear index .
The subscript $\perp$ refers to the transverse $(x,y)$ plane.
We define $\Delta n=n_2|E|^2$ as the nonlinear refractive index.\\
What is remarkable about this equation is that it is similar to the Gross-Pitaevskii equation (GPE), which describes the evolution the wavefunction $\Psi$ of a weakly-interacting Bose-Einstein condensate:
\begin{equation}
    i\hbar\dfrac{\partial\Psi(\textbf{r},t)}{\partial t}=\left(-\dfrac{\hbar^2}{2m}\nabla^2+\mathcal{V}(\textbf{r})+g|\Psi(\textbf{r},t)|^2\right)\Psi(\textbf{r},t) \text{ ,}
    \label{gpe}
\end{equation}
where $\mathcal{V}$ is the trapping potential, m is the bosonic mass and $g$ is the interaction parameter, which is positive for a stable condensate.\\
A major difference between the equations \eqref{nlse} and \eqref{gpe} is that in the NLSE, the time-derivative is replaced by a spatial derivative in the direction of propagation $z$.
To map the NLSE onto the GPE, we define an effective time $\tau=z/c$.
This space-time mapping means that each transverse plane inside the non-linear medium is formally analogous to a 2D Bose gas of photons after the corresponding effective time of evolution $\tau$.
Since the $z$ dimension acts as an effective time dimension, this configuration is referred as 2D+1 geometry. \\
The comparison between the NLSE and the GPE yields  expressions for the effective photon mass $\overline{m}=\hbar k/c$ and for the interaction term $\overline{g}=-\hbar \Delta n$. 
In our case, the stability condition $\overline{g}>0$ corresponds to $\Delta n<0$ (i.e. self defocusing regime).
One can notice that there is no trapping potential term in Eq. \eqref{nlse}, as fluids of light do not need to be held in a trap. 
\begin{figure}
    \includegraphics[width=\columnwidth]{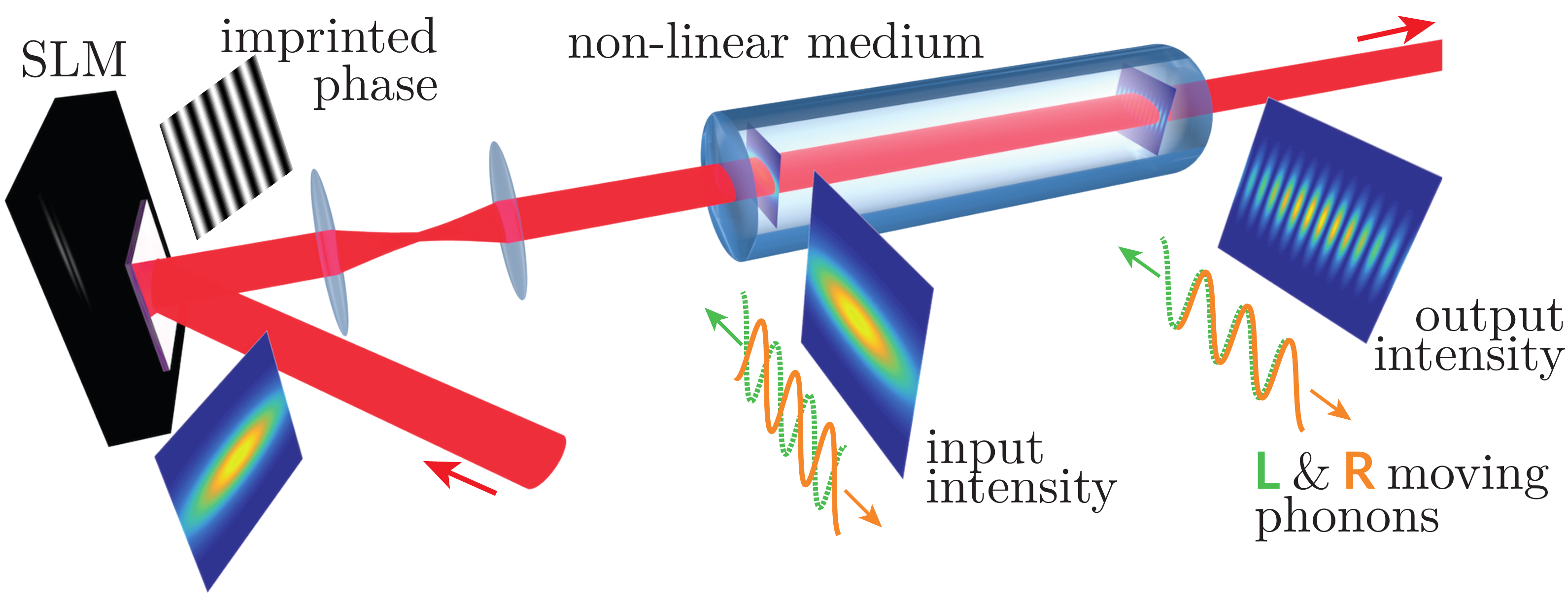}
    \caption{Principle of the experiment and simplified setup. 
    A 780~nm laser beam is elongated in the $x$-direction and sent on a spatial light modulator (SLM).
    The SLM displays a vertical grating, which imprints a sinusoidal phase modulation of wavevector $k_x$. 
    The SLM plane is imaged at the input of the 6.8 cm rubidium vapor cell. 
    This phase modulation creates two counter-propagating left (L) and right (R) phonons at +$k_x$ and -$k_x$ represented in green and orange. 
    Initially in phase opposition (constant input density), the phonons constructively interfere after some effective time $\tau=z/c$, giving a maximum of density contrast.
    The output plane of the non-linear medium (a rubidium vapor cell) is imaged on a camera to study the fringes contrast at z=L.}
    \label{spacetime}
\end{figure}

In a weakly-interacting BEC the excitation spectrum is given by the Bogoliubov dispersion relation $ \Omega_B(\textbf{k}_x)$ \cite{pitaevskii2016bose}, which can be rewritten for a photon fluid as \cite{carusotto2014superfluid}:
\begin{equation}
    \Omega_B(\textbf{k}_x)=\sqrt{(\dfrac{k_x^2}{2k_0})^2+\Delta n k_x^2} \text{ ,}
    \label{bogo_formula}
\end{equation}
Within space-time mapping, $\Omega_B$ has units of inverse of a length.
Eq.~\eqref{bogo_formula} shows two regimes of dispersion, whose transition happens around $k_x=k_0\sqrt{|\Delta n|}=1/\xi$ with $\xi$ is the healing length.
For $k_x\xi<1$, the dispersion is linear and Bogoliubov excitations present a phonon-like behavior: $\Omega_B \approx k_x\Delta n$.
For $k_x\xi>1/$, the dispersion becomes quadratic: $\Omega_B \approx k_x^2/2k_0$. In this regime, excitations behave like massive free particles. 


Bragg spectroscopy in atomic BEC relies on counting the number of scattered atoms as a function of the frequency difference between two Bragg beams.
A variant of this configuration has been presented in \cite{shammass2012} and relies on short Bragg pulses at two symmetrically tilted angles to imprint a phase pattern on a BEC at time $t=0$.
The short pulse results in a broad frequency content, which ensures the creation of counter-propagating phonons at wave-vectors $+k_x$ and $-k_x$, which corresponds to a standing wave in the BEC density.
The density perturbation after a time $t$, defined by $\delta n(t)=|\psi(t)|^2-|\psi(t=0)|^2$, is given by (\ref{rdensity_shammass}):
\begin{equation}
    \delta n(t) =U  S_0(k) \text{cos}(\textbf{k.r})\text{sin}(\omega t) \text{ ,}
    \label{rdensity_shammass}
\end{equation}
where U is a constant quantifying the excitation strength and $S_0(k)$ is the zero-temperature static structure factor.
In \cite{shammass2012}, the authors measured the Fourier transform $\rho_k$ of $\delta n(t)$, (which also oscillates at $\omega(k_x)$)  at different times. 
They extracted the Bogolioubov pulsation from the zero-crossings of $\rho_k$, and the zero-temperature static structure factor from the extrema.

To design an analogue technique for paraxial fluids of light, we had to make two major modifications.
First, we only have access to  one value of $t$ which is given,  in our analogy, by the length of the non-linear medium. 
Therefore, instead of probing the density perturbation as function of time, we probe it as function of $k$ at fixed effective time $\tau=L/c$ 
From Eq.~\eqref{rdensity_shammass}, it is clear that we can still obtain the dispersion relation from the minima of $\delta n$ and the structure factor from the maxima of $\delta n$.
We explain latter in the text how the constant factor $U$ can be canceled with a normalization procedure using a non-interacting gas.

The second important change is on the creation of the phase modulation.
Since we use a fluid of light, we can directly imprint a phase on the laser beam with a spatial light modulator (SLM) and image it on the input plane of the medium.
By imposing a sinusoidal phase pattern on the SLM with a given wavelength and a given depth, we create two left and right propagating phonons (see~Fig.~\ref{spacetime}) with the exact same characteristics as in short Bragg pulse spectroscopy.
This is in fact a general strength of paraxial fluids of light, since any phase modulation (analogous to any short external potential) can be applied on the initial state of our system.
%
%
\begin{figure}
\includegraphics[width=\columnwidth]{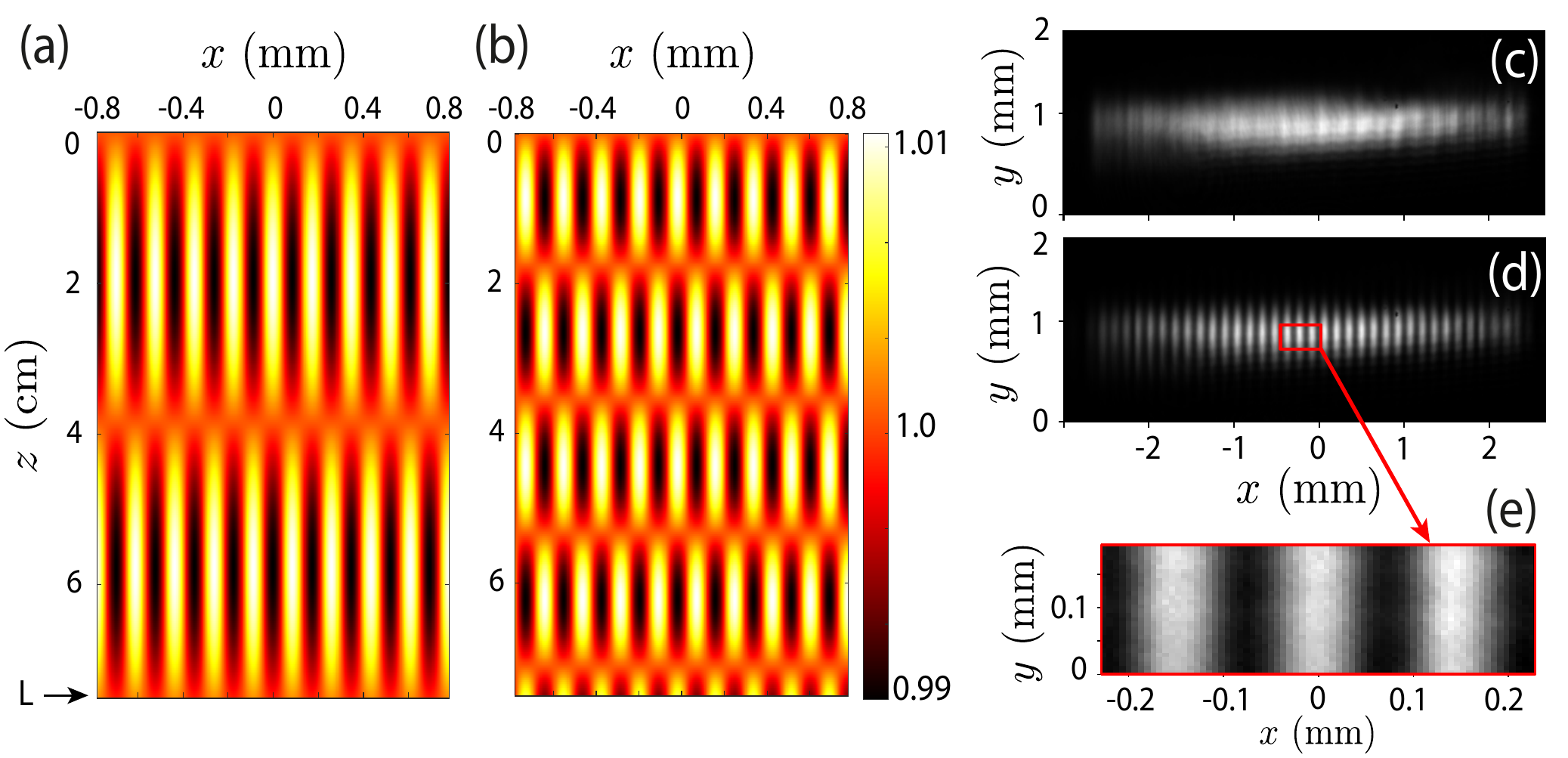}
\caption{(a) and (b) show the simulated density modulation for medium of 7.5 cm and a transverse wave-vector of $k_x$=3.6$\times$ 10$^{4}$~m$^{-1}$, with $\Delta$n=0 (a) and $\Delta$n=2.1$\times$10$^{-5}$ (b). 
(c) and (d) show experimental images of the density obtained respectively for $k_x$=27.8~mm$^{-1}$ (minimum of contrast) and $k_x$=43.2~mm$^{-1}$ (maximum of contrast) in the non-interacting case ($\Delta n$=0). 
These images are taken with a large modulation depth for illustration.
(e) gives a inset of (d).
}
\label{beam}
\end{figure}

Propagation of the density perturbation $\delta n(z)$, calculated for a 1D+1 fluid of light with sinusoidal phase modulation at $k_x$, is shown in Fig.~\ref{beam}~(a) and \ref{beam}~(b).
Interference fringes along the transverse axis $x$ with wave-vector $k_x$ and fringes along the propagation axis $z$ (i.e. effective time $\tau$) with frequency $\omega(k_x)$ can be observed.
Fig.~\ref{beam}(a) presents the simulated pattern in the non-interacting case while Fig~\ref{beam}(b) includes repulsive interactions given by $|\Delta$n$|$=2.1$\times $10$^{-5}$.
The reduction of fringe period along $z$ reflects the difference between the free-particle dispersion (a) and the Bogoliubov dispersion (b) due to the interaction energy.
As mentioned earlier, we cannot directly observe the time evolution in our paraxial fluid of light, since we can only see it at the effective times $\tau$=0 and $\tau$=L/c.
However, we can measure the density contrast at the end of the medium ($z=L$) and identify the contrast minima as function of $k_x$. 
For a given modulation $k_x$, if we define the contrast at $t=L/c$ as  $C =  (n_{max}-n_{min})/(n_{max}+n_{min})$, we can show using Eq.~\eqref{rdensity_shammass} that $C=U  S_0(k)\text{sin}(\omega t)$.
Minima occurs when 
$    \omega (k_x)=p\dfrac{\pi}{L},$
where p is an integer.
The knowledge of the successive minimum locations in $k_x$ will therefore give the dispersion relation, since the corresponding value of $\omega$ is known.

\begin{figure}
\includegraphics[width=8.5cm]{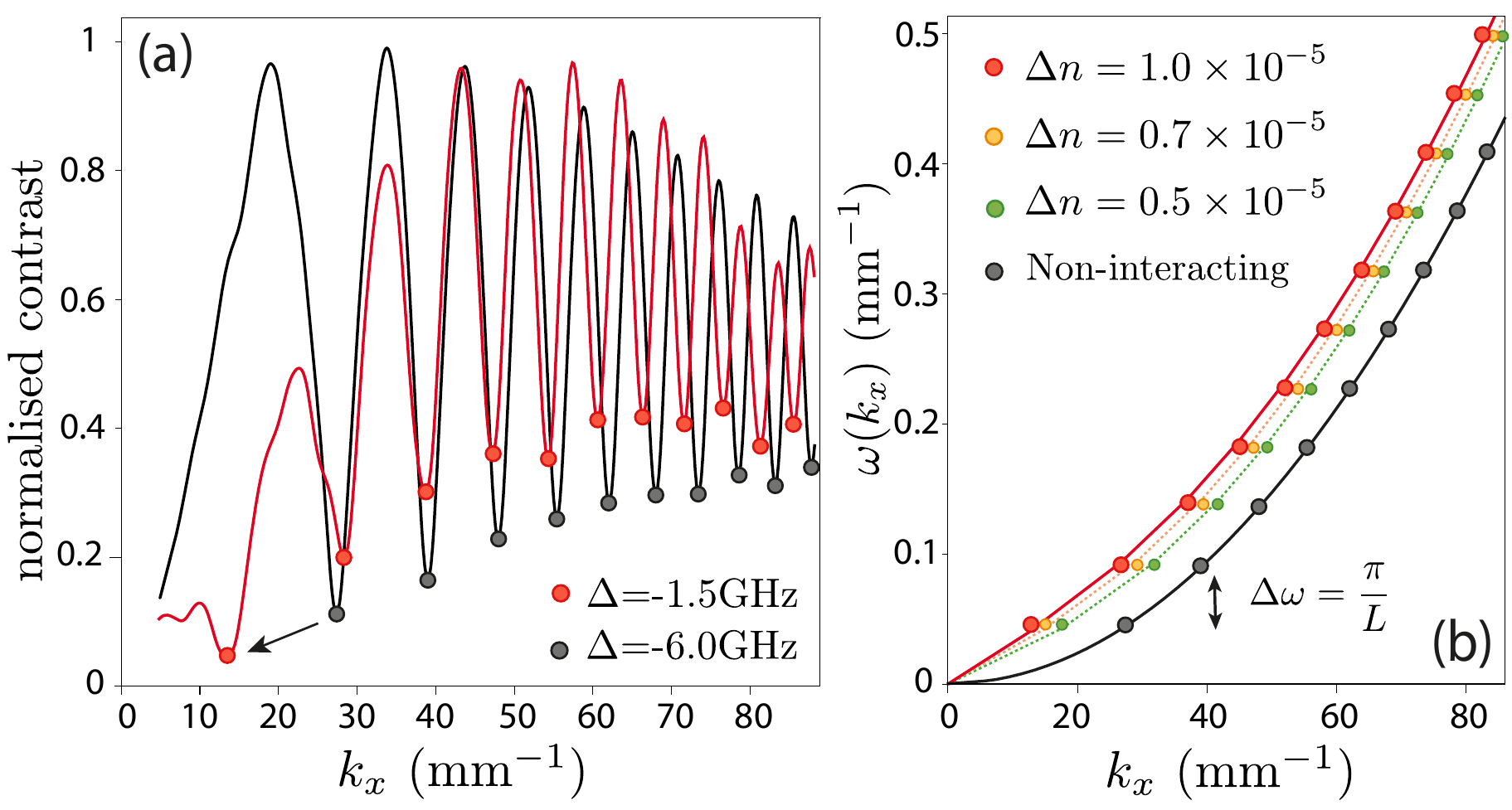}
\caption{(a) Contrast results (smoothed and normalized) for the reference far-off resonance case (grey) with $\Delta$n=0 and for a close-to-resonance set (red) with an input power of 90mW, a cell temperature of 128$^{\circ}$C and an absorption of 50\%. 
The shift of the minima of contrast toward the smaller $k_x$ for the -1.5 GHz set, showed by the grey arrow, is an evidence of the nonlinear effect taking place.
(b) Dispersion $\omega(k_x)$ extracted from the minima of contrast of the reference far-off-resonance set, and for three data sets with decreasing detunings $\Delta$=-2.5 GHz, $\Delta$=-2 GHz and $\Delta$=-1.5 GHz with a laser power increased to 105 mW.
The dots show the extracted experimental values $\omega(k_x)=p\pi/L$ for the p$^{th}$ minimum and the full lines the fits of the Bogolioubov dispersion.
The resulting values of $\Delta n$ are indicated in the legend.}
\label{results_disp}
\end{figure}
To implement this procedure experimentally, we use a 780~nm diode-laser for which we  can finely adjust the detuning with respect to the D$_2$ resonance line of $^{87}$Rb. 
The beam is elongated with two cylindrical lenses to cover the entire surface of the SLM along the $x$ direction.
A sinusoidal phase modulation along $x$ is imprinted using a SLM and is kept sufficiently weak that the resulting intensity modulation does not locally modify the nonlinear index $\Delta n$.
The phase modulation depth is limited to 10\%, to remain within the Bogoliubov perturbative approximation.
For a given $k_x$, the phase applied on the SLM is $\phi(x)=0.1\cos(k_x.x)$.
In order to eliminate the unmodulated reflection on the SLM, we superpose a vertical grating to the horizontal sinusoidal one and select only the first vertical order in the Fourier plane.
The SLM is imaged at the entrance of the non-linear medium with a demagnification factor of three to increase its resolution.
The beam waists at the medium entrance are $w_x$=0.15 mm, $w_y$=1.5 mm.
We consider a local density approximation in order to compare the experimental data with the 1-dimensional simulations.
We verified that the fringe wavelengths were in agreement down to 1\% with the $k_x$ imposed on the SLM.

As a non-linear medium we used a 6.8~cm rubidium cell containing a natural mixture of 28\% of $^{87}$Rb and 72\% of $^{85}$Rb. 
The interaction parameter $\Delta n$ is tuned by adjusting the cell temperature, hence the atomic density in the cell \cite{glorieux2011quantum,agha2011time}, the laser detuning and the laser power. 
One should be careful that the linear absorption $\alpha$ increases as well when going closer to resonance.
We estimate a maximum level of absorption to not perturb the model to be $\alpha=13~\text{cm}^{-1}$ (60\% transmission), and make sure to work under this value.
The output plane of the cell is then imaged on a camera. 
Typical beam picture obtained for the non-interacting case ($\Delta n$=0) are shown on Fig.~\ref{beam}(c) and \ref{beam}(d) for $k_x$=27.8~mm$^{-1}$  (contrast minimum) and $k_x$=43.2~mm$^{-1}$ (contrast maximum). 
This non-interacting case is obtained experimentally by setting a large detuning ($\Delta$=-6~GHz) from the $^{87}$Rb $F=2$ to $F'$ transition.



We then took 475 images of the cell output with modulations ranging from $k_x$=5~mm$^{-1}$ to $k_x$=100~mm$^{-1}$ with a step $dk_x$=0.2~mm$^{-1}$.
To measure the density modulation $\delta n$, we normalize the images by a reference taken without phase modulation and then select a central window of the fluid (see Fig.\ref{beam}(e)).
After integration of the intensity along the $y$ (vertical) axis, we calculate the contrast which is directly proportional to $\delta n(\tau=L/c)$.
In figure \ref{results_disp}(a), we plot the normalized contrast as a function of $k_x$ and highlight the minima positions for a non-interacting and a weakly interacting ($\Delta$=-1.5 GHz) fluid of light.
For the latter case, we see a clear shift of the contrast minima  toward smaller values of $k_x$, which indicates the presence of interaction.

In Fig.~\ref{results_disp}(b), we plot the $k_x$ positions of the contrast minima (marked in Fig.~\ref{results_disp}(a) and shifted vertically of $\pi/L$ as explained previously) using four different detunings ($\Delta$=-2.5 GHz, $\Delta$=-2 GHz and $\Delta$=-1.5 GHz with a laser power of 105 mW).
The fits using  Eq.~\eqref{bogo_formula} to extract the value of $\Delta n$ which quantify the interactions are given in solid lines.
The dots are the experimental points and the full lines are fits to the Bogolioubov relation. 
The far-off resonance set matches exactly the quadratic dispersion $\omega=k_x^2/2k_0$, which corresponds to $\Delta n=0$ with an uncertainty of $2\times 10 ^{-7}$ for this set.
Then, for the three weakly interacting sets with detunings $\Delta$=-2.5~GHz (green), $\Delta$=-2~GHz (yellow), $\Delta$=-1.5~GHz (red), the fits of the Bogolioubov dispersion relations give the respective values of $\Delta n=5\times 10 ^{-6}$, $\Delta n=7 \times 10^{-6}$ and $\Delta n=10 \times 10 ^{-6}$.

We can evaluate the sensitivity of this method for resolving weak interaction (small $|\Delta n|$).
The energy offset accumulated in the linear part of the dispersion translates into an energy shift at large $k_x$. The dispersion curve with $\Delta n \neq$0 is vertically shifted relative to the non-interacting one ($\Delta n=0$).
We can calculate this shift at high $k_x$ as:
\begin{equation}
    \Omega_B(k_x)-\omega_{lin}(k_x)=\sqrt{k_x^2|\Delta n| + (\frac{k_x^2}{2k_0})^2}- \frac{k_x^2}{2k_0} \approx k_0|\Delta n|
    \label{high_k_disp} \text{ .}
\end{equation}
The value of $\Delta$n can then be directly obtained by the difference with the non-interacting reference at high $k_x$. We verified the validity of this method with the 3 datasets presented  on Fig.~\ref{results_disp}(b).
Knowing the uncertainty on $k_x$ to be $dk_x$=0.2~mm$^{-1}$,  it is possible to estimate the smallest nonlinear index value achievable with this technique to be $\Delta n = 2\times10^{-7}$. 
This is more than an order of magnitude better than previous techniques using a group velocity measurement \cite{fontaine2018}.

\begin{figure}
\includegraphics[width=8.5cm]{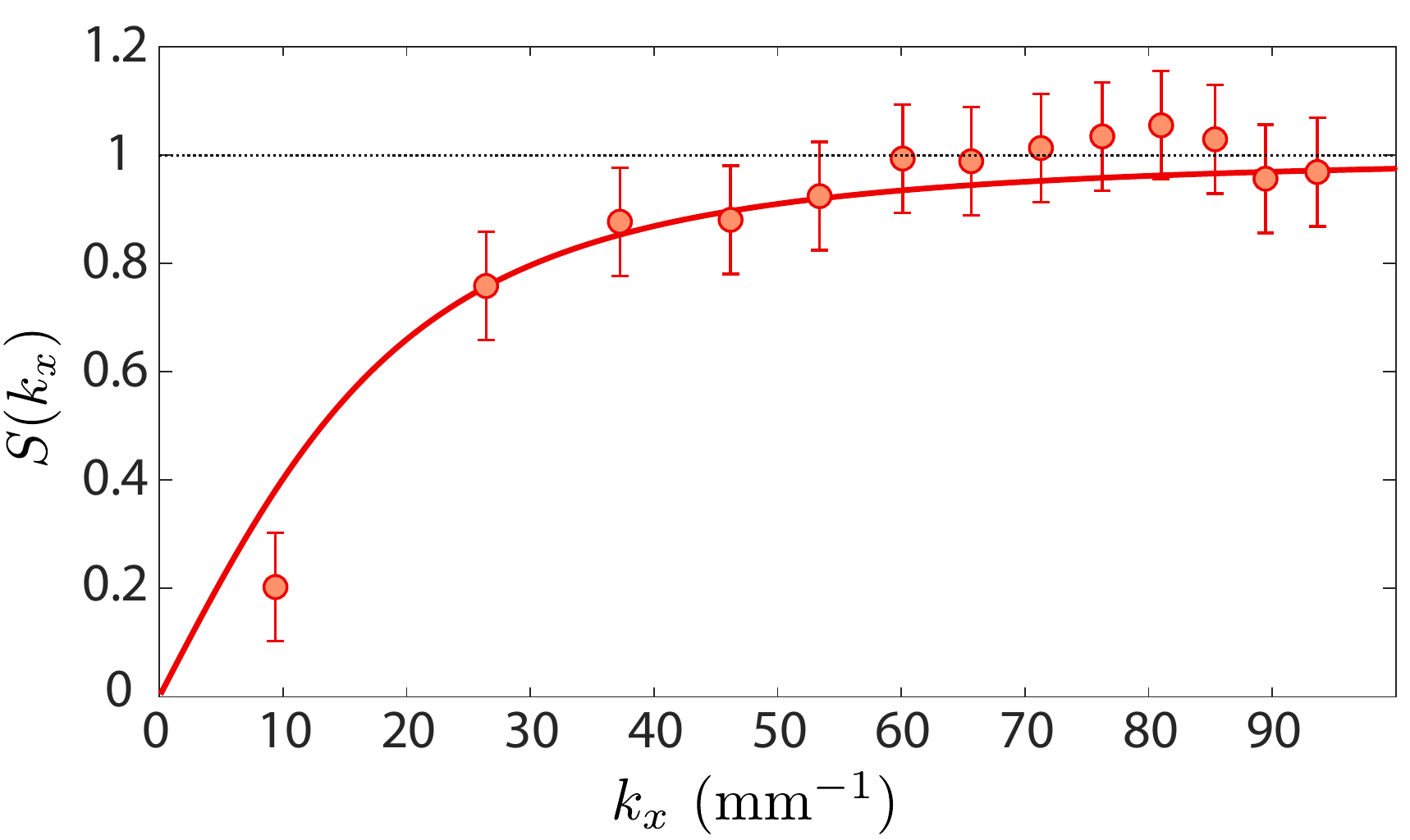}
\caption{Zero temperature static structure factor measurement. 
The experimental conditions are the same as green curve of figure \ref{results_disp}, but at a temperature 20 degrees lower (108$^{\circ}$C), leading to $\Delta n=2\times 10^{-6}$.
The solid red line is the model given by \ref{skratio}. 
Dashed line at $S(k_x)=1$ is the structure factor of a non-interacting gas (coherent state).}
\label{results_sk}
\end{figure}

Finally, we present a measurement of the static structure factor $S(k_x)$ in a fluid of light.
The structure factor is the spatial noise spectrum (normalized to 1 for a coherent state).
From \eqref{rdensity_shammass}, we explained how to isolate $US(k_x)$ using the the contrast maxima values.
In order to remove the $U$ dependance, we measure $US(k_x)$ for the non-interacting case.
In this regime it is known that $S(k_x)$ must be equal to 1 at all $k_x$ since the beam here is a spatially coherent state. 
This allows us to measure $U$ which actually depends on $k_x$ \cite{pitaevskii2016bose}.
$U$ is determined by the phase imprinting efficiency of the SLM, the modulation transfer function of the optical system and the depth of phase modulation.
These three parameters are not modified while changing $\Delta n$, therefore the $S(k_x)$ for the interacting case is obtained by dividing the contrast maxima values for $\Delta n \neq 0$ by the calibrated value of $U$.
The results are presented in Fig.~\ref{results_sk} for a weakly interacting fluid ($\Delta n=2\times 10^{-6}$) as well as the Feynman relation \cite{nozieres2018theory,griffin1993excitations,feynman1954atomic}:
\begin{equation}
  S(k_x)= \frac{k_x^2/2k_0}{ \Omega_B(k_x)}.
  \label{skratio}
\end{equation}
Our experimental data (dots) show a good agreement with the Feynman relation (solid line) with no free parameters, since the value of $\Delta n$ is obtained using \eqref{high_k_disp}.
The structure factor can be interpreted as the rate of producing excitations at momentum $k_x$ in a Bose gas \cite{PhysRevLett.88.120407}.
Fig.~\ref{results_sk} clearly shows that the rate of Bragg excitations is highly reduced at low $k_x$.
This can be explained by quantum depletion in our fluid of light \cite{PhysRevLett.83.2876}.
Indeed, observing $S(k_x)<1$ at low $k_x$ is a consequence of the creation of correlated pairs at $+k_x$ and $-k_x$ which minimize the total energy of the system, known as quantum depletion \cite{PhysRevLett.83.2876}. 
In our experimental configuration the number of points where this effect is highly visible (when $k_x<1/\xi$) is limited by the cell length $L$.
Indeed, if we require that the first contrast minimum occurs at  $k_x=1/\xi=k_0\sqrt{|\Delta n|}$, then the cell length is given by $\L=\dfrac{2\pi}{\sqrt{5}k_0|\Delta n|}$. 
In order to increase the number of point below  $k_x=1/\xi$ further experiments could increase the cell length but this will be ultimately limited by linear absorption.\\

\section{Conclusion}
In this work, we have implemented Bragg spectroscopy in a paraxial fluid of light. 
This technique has proved to be an essential tool to study ultracold atomic BEC and was missing in fluid of light platforms.
We show that our implementation is robust and highly sensitive since it allows to measure the interactions an order of magnitude weaker than previously reported.
Importantly, we present a measurement of the zero temperature static structure factor and show a good agreement with the Feynman relation.
This measurement of the structure factor also reveals the presence of quantum  depletion,  consisting  of  pair-correlated particles, in a paraxial fluid of light.
These results open the way to the measurement of  the Tan's contact and the observation of beyond mean field effects in photon fluids.


\begin{acknowledgments}
This work has received funding from the French ANR grants (C-FLigHT 138678 and "Quantum Fluids of Light" ANR-16-CE30-0021) and from the European Union Horizon 2020 Research and Innovation Program under Grant Agreement No. 820392 (PhoQuS). Q. G. and A. B. thank the Institut Universitaire de France (IUF).
The authors thank Antonine Rochet for support with 3D figure.
\end{acknowledgments}

\appendix

\nocite{*}

\bibliography{braggfluid}

\end{document}